\documentclass[floats,nofootinbib,prb,twocolumn]{revtex4}%
\usepackage{amsfonts}
\usepackage{amsmath}
\usepackage{amssymb}
\usepackage{graphicx}%
\begin{document}
\title{{\Large Environmentally induced Quantum Dynamical Phase Transition in the spin
swapping operation.}}
\author{Gonzalo A. \'{A}lvarez}
\affiliation{Facultad de Matem\'{a}tica, Astronom{\'{\i}}a y F{\'{\i}}sica, Universidad
Nacional de C\'{o}rdoba, Ciudad Universitaria, 5000, C\'{o}rdoba, Argentina}
\author{Ernesto P. Danieli}
\affiliation{Facultad de Matem\'{a}tica, Astronom{\'{\i}}a y F{\'{\i}}sica, Universidad
Nacional de C\'{o}rdoba, Ciudad Universitaria, 5000, C\'{o}rdoba, Argentina}
\author{Patricia R. Levstein}
\email{patricia@famaf.unc.edu.ar}
\affiliation{Facultad de Matem\'{a}tica, Astronom{\'{\i}}a y F{\'{\i}}sica, Universidad
Nacional de C\'{o}rdoba, Ciudad Universitaria, 5000, C\'{o}rdoba, Argentina}
\author{Horacio M. Pastawski}
\email{horacio@famaf.unc.edu.ar}
\affiliation{Facultad de Matem\'{a}tica, Astronom{\'{\i}}a y F{\'{\i}}sica, Universidad
Nacional de C\'{o}rdoba, Ciudad Universitaria, 5000, C\'{o}rdoba, Argentina}

\begin{abstract}
Quantum Information Processing relies on coherent quantum dynamics\ for a
precise control of its basic operations. A swapping gate in a two-spin system
exchanges the degenerate states $\left\vert \uparrow,\downarrow\right\rangle $
and $\left\vert \downarrow,\uparrow\right\rangle $. In NMR, this is achieved
turning on and off the spin-spin interaction $b=\Delta E$ that splits the
energy levels and induces an oscillation with a natural frequency $\Delta
E/\hslash$. Interaction of strength $\hbar/\tau_{\mathrm{SE}}$, with an
environment of neighboring spins, degrades this oscillation within
a\ decoherence time scale $\tau_{\phi}$. While the experimental frequency
$\omega$ and decoherence time $\tau_{\phi}$ were expected to be roughly
proportional to $b/\hbar$ and $\tau_{\mathrm{SE}}$ respectively, we present
here experiments that show drastic deviations in both $\omega$ and $\tau
_{\phi}$. By solving the many spin dynamics, we prove that the swapping regime
is restricted to $\Delta E\tau_{\mathrm{SE}}\gtrsim\hslash$. Beyond a critical
interaction with the environment the swapping freezes and the decoherence rate
drops as $1/\tau_{\phi}^{{}}\propto\left(  b/\hbar\right)  _{{}}^{2}%
\tau_{\mathrm{SE}}^{{}}$. The transition between quantum dynamical phases
occurs when $\omega\propto\sqrt{(b/\hslash)_{{}}^{2}-\left(  k/\tau
_{\mathrm{SE}}^{{}}\right)  ^{2}}$ becomes imaginary, resembling an overdamped
classical oscillator. Here, $0\leq k^{2}\leq1$ depends only on the anisotropy
of the system-environment interaction, being $0$ for isotropic and $1$ for XY
interactions. This critical onset of a phase dominated by the Quantum Zeno
effect opens up new opportunities for controlling quantum dynamics.

\end{abstract}
\startpage{1}
\maketitle

\section{INTRODUCTION}

Experiments on quantum information processing \cite{BD2000}\ involve atoms in
optical traps \cite{Myatt2000}, superconducting circuits \cite{urbina2002} and
nuclear spins \cite{cory2003,Awschalom2003} among others. Typically, the
system to be manipulated interacts with an environment
\cite{Myatt2000,Gurvitz2003,Zurek2003,Barret2004} that perturbs it, smoothly
degrading its quantum dynamics with a decoherence rate, $1/\tau_{\phi}$,
proportional to the system-environment (SE) interaction $\hbar/\tau
_{\mathrm{SE}}$. Strikingly,\ there are conditions where the decoherence
rate\ can become perturbation independent.\cite{PhysicaA} This phenomenon is
interpreted \cite{JalPas,Beenakker,CookPasJal} as the onset of a Lyapunov
phase, where $1/\tau_{\phi}=\min\left[  1/\tau_{\mathrm{SE}},\lambda\right]  $
is controlled by the system's own complexity $\lambda$. Describing such a
transition, requires expressing the observables (outputs) in terms of the
controlled parameters and interactions (inputs) beyond the perturbation
theory. We are going to show that this is also the case of the simple swapping
gate, an essential building block for quantum information processing, where
puzzling experiments require a substantially improved description. While the
swapping operation was recently addressed in the field of NMR in liquids
\cite{Madi-Ernst, Freeman} with a focus on quantum computation, the pioneer
experiments were performed in solid state NMR \cite{MKBE} by M\"{u}ller,
Kumar, Baumann and Ernst (MKBE). They obtained a swapping frequency $\omega$
determined by a two-spin dipolar interaction $b$, and a decoherence rate
$1/\left(  2\tau_{\phi}\right)  \equiv R$ that, in their model, was fixed by
interactions with the environment $1/\left(  2\tau_{\mathrm{SE}}\right)  $.
This dynamical description was obtained by solving a generalized Liouville-von
Neumann equation. As usual, the degrees of freedom of the environment were
traced out to yield a quantum master equation.\cite{QME} More recent
experiments, spanning the internal interaction strength \cite{JCP98} hinted
that there is a critical value of this interaction when a drastic change in
the behavior of the swapping frequency and relaxation rates occurs. Since this
is not predicted by the standard approximations in the quantum master
equation,\cite{MKBE} this motivates to deepen into the physics of the phenomenon.

In this paper, we present a set of $^{13}$C-$^{1}$H cross-polarization NMR
data, swept over a wide range of a control parameter (the ratio between
internal interactions and SE interaction strengths). These results clearly
show that the transition\textit{ }between the two expected dynamical regimes
for the $^{13}$C polarization, an oscillating regime and an over-damped
regime, is not a smooth cross-over. Indeed, it has the characteristics of
critical phenomena where a divergence of the oscillation period at a given
critical strength of the control parameter is indicative of the nonanalyticity
of this observable.\cite{Horsthemke-Lefever, sachdev} The data are interpreted
by solving the swapping dynamics between two coupled spins (qubits)
interacting with a spin bath. With this purpose the environment is represented
as a stroboscopic process. With certain probability, this instantaneous
interaction interrupts the system evolution through measurements and/or
injections. This simple picture, emerging naturally \cite{GLBE1,GLBE2} from
the quantum theory of irreversible processes in the Keldysh
formalism,\cite{Keldysh,Keldysh2} enables us to distinguish the interaction
parts that lead to dissipation from those giving pure decoherence. Within this
picture, the overdamped regime arises because of the quantum Zeno
effect,\cite{Misra,Usaj,PascazioQZSub2002} i.e. environment \textquotedblleft
measures\textquotedblright\ the system so frequently that prevents evolution.
The analytical solution confirms that there is a critical value of the control
parameter where a bifurcation occurs. This is associated with the switch among
dynamical regimes: the \textit{swapping} \textit{phase }and the \textit{Zeno
phase}. In consequence, we call this phenomenon a \textit{Quantum Dynamical
Phase Transition}.

\section{\textbf{EXPERIMENTAL EVIDENCE}}

\begin{figure*}
[th]
\begin{center}
\includegraphics[
height=4.872in,
width=6.636in
]%
{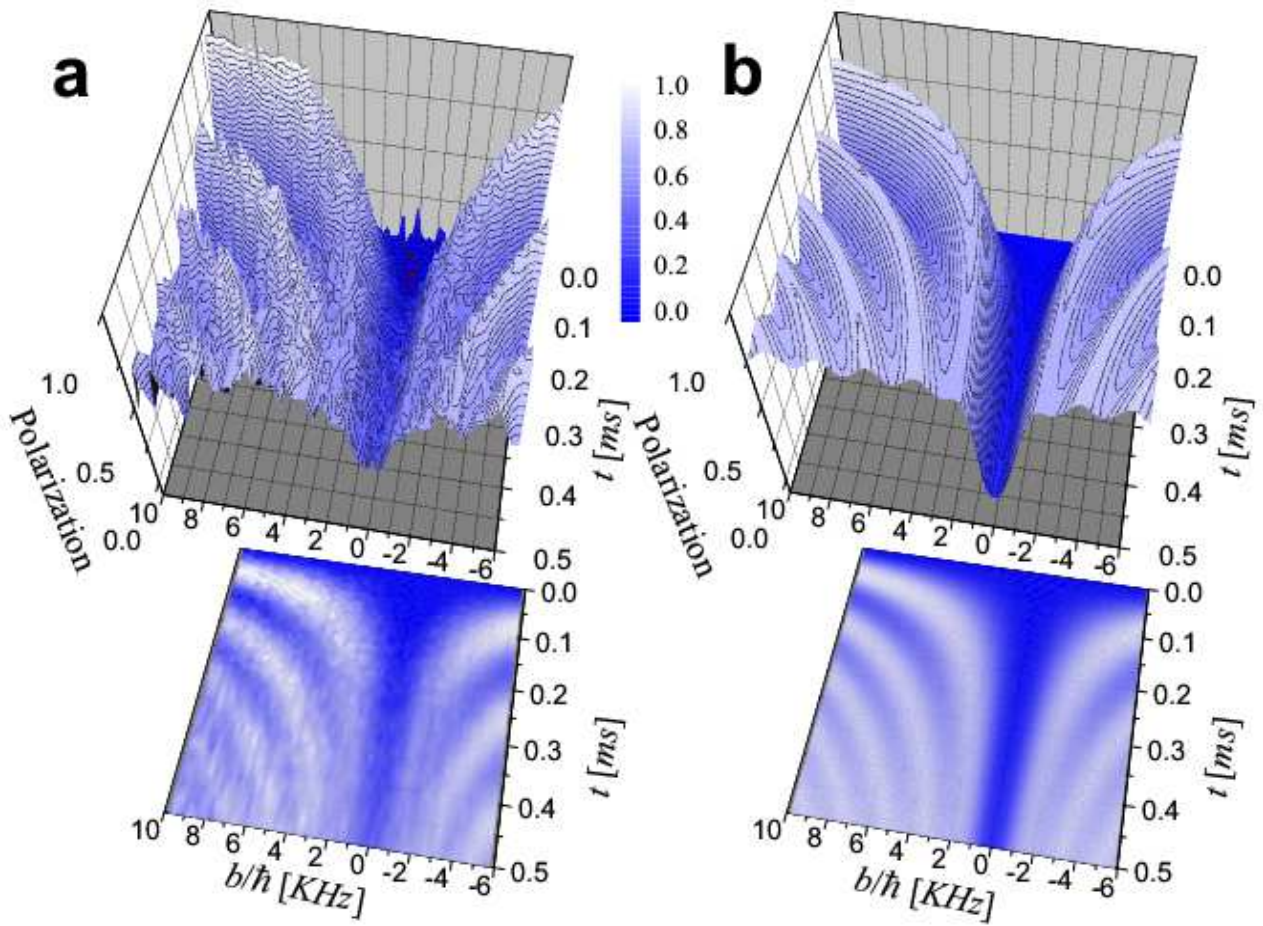}%
\caption{(Color online) Spin swapping dynamics in $^{13}$C-$^{1}$H.
(\textbf{a}): Experimental $^{13}$C polarization in Fe(C$_{5}$H$_{5}$)$_{2}$
as a function of the contact time $t$ and spin-spin coupling $b(\theta)$.
(\textbf{b}): Numerical simulations of the $^{13}$C polarization obtained from
Eq. (\ref{PolCST}) for different values of $b$, a dipolar system-environment
interaction $\left(  \left\vert \alpha/\beta\right\vert =2\right)  $ and a
constant value for $\tau_{\mathrm{SE}}$ ($\tau_{\mathrm{SE}}=0.275~\mathrm{ms}%
$) obtained by fitting the experimental data in the regime where the MKBE
expression is valid. Projection plots in the $b-t$ plane show a canyon where
the oscillation period diverges indicating a Quantum Dynamical Phase
Transition.}%
\label{Figteoexp3D}%
\end{center}
\end{figure*}

Our cross polarization experiments exploit the fact that in polycrystalline
ferrocene Fe(C$_{5}$H$_{5}$)$_{2}$, one can select a pair of interacting
spins, i.e. a $^{13}$C and its directly bonded $^{1}$H, arising on a molecule
with a particular orientation. This is because the cyclopentadienyl rings
perform fast thermal rotations ($\approx$ $\mathrm{ps}$) around the five-fold
symmetry axis, leading to a time averaged $^{13}$C-$^{1}$H interaction. The
new dipolar constant depends \cite{slichter} only on the angle $\theta$
between the molecular axis and the external magnetic field $B_{0}$ and the
angles between the internuclear vectors and the rotating axis, which in this
case are 90$^{\circ}$. Thus, the effective coupling constant is%

\begin{equation}
b=\frac{1}{2}\frac{\mu_{0}\ \gamma_{\mathrm{H}}\,\gamma_{\mathrm{C}}%
\,\hbar^{2}}{4\pi r_{\mathrm{HC}}^{3}}\frac{\left\langle 3\cos^{2}%
\theta-1\right\rangle }{2},\label{b}%
\end{equation}
where $\gamma$'s are the gyromagnetic factors and $r_{\mathrm{HC}}^{{}}$ the
internuclear distance. Notice that $b(\theta)$ cancels out at the magic angle
$\theta_{m}\simeq54.74^{\circ}.$ As the chemical shift anisotropy of $^{13}$C
is also averaged by the rotation and also depends on $\theta$ as $\left\langle
3\cos^{2}\theta-1\right\rangle ,$ it is straightforward to assign each
frequency in the $^{13}$C spectrum to a dipolar coupling $b$. Thus, all
possible $b$ values are present in a single polycrystalline spectrum. The
swapping induced by $b$ is turned on during the \textquotedblleft contact
time\textquotedblright\ $t$, when the sample is irradiated with two radio
frequencies fulfilling the Hartmann-Hahn condition.\cite{slichter}
Experimental details have been given elsewhere.\cite{JCP98} At $t=0,$ there is
no polarization at $^{13}$C while the $^{1}$H system is polarized. The
polarization is transferred forth and back in the $^{13}$C-$^{1}$H pairs while
the other protons inject polarization into these pairs. In Fig.
\ref{Figteoexp3D}a, we show the raw experimental data of $^{13}$C polarization
as a function of the contact time and $b(\theta)$. The polarizations have been
normalized to their respective values at the maximum contact time
($3~\mathrm{ms}$) for each $\theta$ when it saturates. It can be appreciated
in the figure that the oscillation frequency is roughly proportional to
$\left\vert b\right\vert ,$ showing that this is the dominant interaction in
the dynamics. This is consistent with the fact that\ the next $^{13}$C-$^{1}$H
coupling strength with a non-directly bonded proton is roughly $b/8$ and, as
all the intramolecular interactions, also scales with the angular factor
$\left\langle 3\cos^{2}\theta-1\right\rangle $.%

A noticeable feature in these experimental data is the\ presence of a
\textquotedblleft canyon\textquotedblright,\ in the region $\left\vert
b\right\vert <2~$\textrm{kHz}, where oscillations (swapping) disappear. The
white hyperbolic stripes in the contour plot at the bottom evidence a swapping
period $2\pi/\omega$ that diverges for a non-zero critical interaction. This
divergence is the signature of a critical behavior.

The standard procedure to characterize the cross polarization experiment in
ferrocene and similar compounds\cite{JCP98} is derived from the MKBE model.
There the $^{13}$C polarization exchanges with that of its directly bonded
$^{1}$H, which, in turn, interacts isotropically with other protons that
constitute the environment.\cite{MKBE} Their solution is
\begin{equation}
P^{^{\mathrm{MKBE}}}(t)=1-\frac{1}{2}\exp\left[  -\frac{t}{2\tau_{\phi}%
}\right]  -\frac{1}{2}\cos(\omega t)\exp\left[  -\frac{3}{2}\frac{t}%
{2\tau_{\phi}}\right]  ),\label{eq-MKBE}%
\end{equation}
where the decoherence rate becomes determined by the rate of interaction with
the environment $1/\left(  2\tau_{\mathrm{SE}}\right)  \rightarrow1/\left(
2\tau_{\phi}\right)  \equiv R,$ while the swapping frequency is given by the
two-spin dipolar interaction, $b/\hbar\rightarrow\omega$. A dependence of the
inputs $b$ and $\tau_{\mathrm{SE}}$ on $\theta$ should manifest in the
observables $\omega$ and $\tau_{\phi}.$ However, working on a polycrystal,
each $\tau_{\mathrm{SE}}(\theta)$ value involves a cone of orientations of
neighboring molecules and a rough description with single average value for
the SE interaction rate is suitable.

\begin{figure}
[h]
\begin{center}
\includegraphics[
trim=0.000000in 0.000000in 0.000313in 0.000000in,
height=2.86218in,
width=3.1668in
]%
{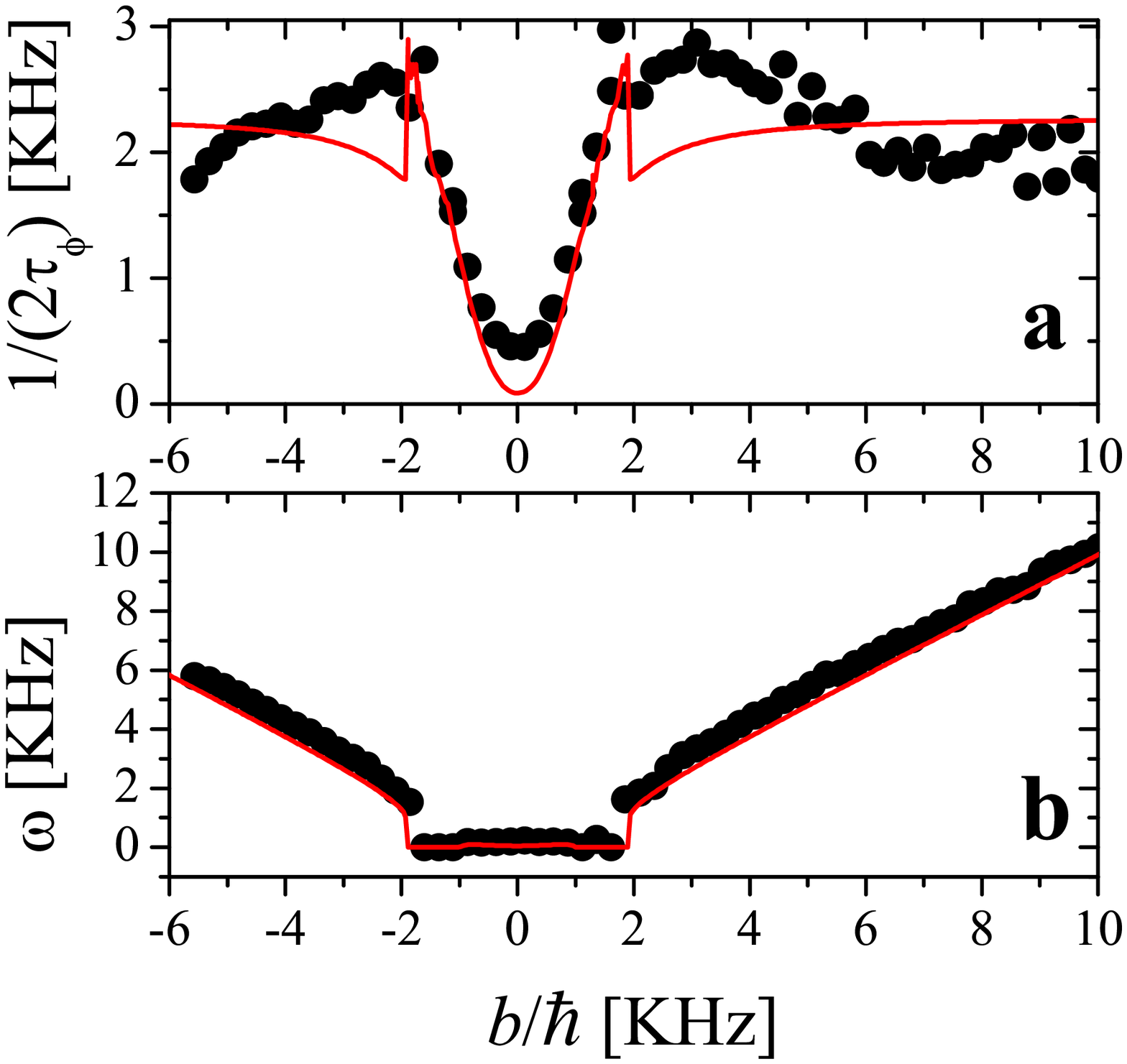}%
\caption{{}(Color online) Decoherence rate 1/(2$\tau_{\phi}$) and frequency
$\omega$ in the spin swapping of a $^{13}$C-$^{1}$H system. Data points are
obtained by fitting cross polarization experiments to the expression
$P^{^{\mathrm{MKBE}}}(t)$. The zero plateau in the frequency and the parabolic
behavior of $1/(2\tau_{\phi})$ in the region $b\tau_{\mathrm{SE}}\ll\hbar$ are
indicative of an over-damped Zeno phase. Solid lines are the prediction of our
model assuming a constant $\tau_{\mathrm{SE}}=0.275\ ms$.}%
\label{Figexp}%
\end{center}
\end{figure}

We have performed non-linear least square fittings of the experimental points
to the equation $P^{^{\mathrm{MKBE}}}(t)$ for the whole $^{13}$C spectra of
ferrocene in steps of $\approx80\mathrm{Hz}$ and contact times ranging from
$2\mathrm{\mu s}$ to $3\mathrm{ms}$. The $1/\left(  2\tau_{\phi}\right)  $ and
$\omega$ parameters obtained from these fits are shown as dots in Fig.
\ref{Figexp}. The proportionality of the frequency with $b$ for orientations
that are far from the magic angle is verified. In this region a weak variation
of $1/\left(  2\tau_{\phi}\right)  $ around $2.2\mathrm{kHz}$ reflects
the\ fact that $1/\left(  2\tau_{\mathrm{SE}}\right)  $ does not depend on
$\theta$. A drawback of this simple characterization is that it tends to
overestimate the width of the canyon because of limitations of the fitting
procedure when Eq. (\ref{eq-MKBE}) is used around the magic angle.%

In spite of the MKBE theoretical prediction, one observes that the frequency
becomes zero abruptly and the relaxation rate suddenly drops with a quadratic
behavior when $b_{\mathrm{c}}\simeq2\mathrm{kHz}$. The minimum of the parabola
occurs at the magic angle, when $b=0.$ Then, all the polarization reaching the
$^{13}$C at this orientation originates from protons outside the molecule.
Then, the rate of $0.5\mathrm{kHz}$ obtained at this minimum constitutes an
experimental estimation of this mechanism. This has to be compared with the
almost constant value of $1/(2\tau_{\mathrm{SE}})=1/(2\tau_{\phi})\simeq$
$2.2\mathrm{kHz}$ observed outside the magic angle neighborhood. This
justifies neglecting the J-coupling and the direct relaxation of the $^{13}$C
polarization through the dipolar interaction with protons outside the
molecule. In the following we describe our stroboscopic model that accounts
for the \textquotedblleft anomalous\textquotedblright\ experimental behavior.

\section{THEORETICAL DESCRIPTION}

\subsection{\textbf{The system}}

Let us consider $M$ coupled $1/2$ spins with a Hamiltonian:%
\begin{equation}
\mathcal{H}=\mathcal{H}_{\mathrm{Z}}+%
{\textstyle\sum\limits_{i<j}}
\left[  a_{ij}I_{i}^{z}I_{j}^{z}+b_{ij}\left(  I_{i}^{+}I_{j}^{-}+I_{i}%
^{-}I_{j}^{+}\right)  /2\right]  , \label{SpinHamiltonian}%
\end{equation}
where $\mathcal{H}_{\mathrm{Z}}=\sum_{i=1}^{M}\hbar\left(  \omega_{\mathrm{L}%
}+\delta\omega_{i}\right)  I_{i}^{z}$ is the Zeeman energy, with a mean Larmor
frequency $\omega_{\mathrm{L}}$. The second term is the spin-spin interaction:
$b_{ij}/a_{ij}=0$ is Ising, and $a_{ij}/b_{ij}=0,1,-2$ gives an $XY$, an
isotropic (Heisenberg) or the truncated dipolar (secular), respectively. This
last case is typical in solid-state NMR experiments \cite{QME} where
$\hbar\delta\omega_{i},a_{ij},b_{ij}\ll\hbar\omega_{\mathrm{L}}$.

In order to describe the experimental system let us take the first $N=2$
spins, $I_{1}$ (a $^{13}$C) and $I_{2}$ (its directly bonded $^{1}$H), as the
\textquotedblleft system\textquotedblright\ where the swapping $\left\vert
\downarrow_{1}\uparrow_{2}\right\rangle \rightleftarrows\left\vert
\uparrow_{1}\downarrow_{2}\right\rangle $ occurs under the action of $b_{12}%
$.\ The other $M-N$ spins (all the other $^{1}$H), with $M\rightarrow\infty$,
are the spin-bath or \textquotedblleft environment\textquotedblright. This
limit enables the application of the Fermi Golden Rule or a more sophisticated
procedure to obtain a meanlife $\tau_{\mathrm{SE}}$ for the system levels. We
will not need much detail for the parameters of the spin bath in Eq.
(\ref{SpinHamiltonian}) except for stating that it is characterized by an
energy scale $d_{\mathrm{B}}$ which leads to a very short correlation time
$\tau_{\mathrm{B}}\simeq\hbar/d_{\mathrm{B}}$.

\subsection{\textbf{Spin}$\leftrightarrow$\textbf{Fermion mapping}}

The spin system can be mapped into a fermion particle system using the
Jordan-Wigner transformation,\cite{spin-fermion,spin-fermion2} $I_{i}%
^{+}=c_{i}^{+}\exp\left\{  \mathrm{i}\pi\sum_{j=1}^{i-1}c_{j}^{+}c_{j}^{{}%
}\right\}  $. Under the experimental conditions, $\delta\omega_{i}=0,$
$a_{12}=0$ and $b_{12}=b,$ the system Hamiltonian becomes:%
\begin{equation}
\mathcal{H}_{\mathrm{S}}=\hbar\omega_{\mathrm{L}}\left(  c_{1}^{+}c_{1}^{{}%
}+c_{2}^{+}c_{2}^{{}}-\mathbf{1}\right)  +b\left(  c_{1}^{+}c_{2}^{{}}%
+c_{2}^{+}c_{1}^{{}}\right)  /2.
\end{equation}
The Jordan-Wigner transformation maps a linear many-body $XY$ spin Hamiltonian
into a system of non-interacting fermions. This leads us to solve a one-body
problem, reducing the dimension of the Hilbert space from $2^{N}$ to $N$
states that represent local excitations.\cite{spin-fermion,spin-fermion2}%
\emph{ }To simplify the presentation, and without loss of generality, we
consider a \textit{single} connection between the system and the spin bath,
$a_{23}=\alpha d_{23}$ and $b_{23}=\beta d_{23}$ with $a_{2j}=b_{2j}=0,$
$j=4\ldots\infty$ and $a_{1j}=b_{1j}=0,$ $j=3\ldots\infty$. Spins $I_{i}$ with
$3\leq i\leq M$ are interacting among them. The SE interaction becomes:%
\begin{equation}
\mathcal{V}=d_{23}\left[  \alpha\left(  c_{2}^{+}c_{2}^{{}}-\tfrac{1}%
{2}\right)  \left(  c_{3}^{+}c_{3}^{{}}-\tfrac{1}{2}\right)  +\beta\left(
c_{2}^{+}c_{3}^{{}}+c_{3}^{+}c_{2}^{{}}\right)  /2\right]  .
\end{equation}
In the first (Ising) term, the first factor \textquotedblleft
measures\textquotedblright\ if there is a particle at site $2$ while the
second \textquotedblleft measures\textquotedblright\ at site$\ 3$. Hence,
polarization at site $2$ is \textquotedblleft detected\textquotedblright\ by
the environment. The hopping term swaps particles between bath and system.

In the experimental initial condition, all spins are polarized with the
exception of $I_{1}$.\cite{MKBE,JCP2003} In the high temperature limit
($\hbar\omega_{\mathrm{L}}/k_{B}T\equiv s\ll1$), the reduced density operator
is $\mathbf{\sigma}\left(  0\right)  =\frac{\hbar}{\mathrm{i}}\mathbf{G}%
^{<}\left(  0\right)  =\left(  \mathbf{1}+sI_{2}^{z}\right)
/\operatorname{Tr}\left\{  \mathbf{1}\right\}  $ which under the Jordan-Wigner
transformation becomes $\frac{\left(  1-s/2\right)  }{\operatorname{Tr}%
\left\{  \mathbf{1}\right\}  }\mathbf{1}+\frac{s}{\operatorname{Tr}\left\{
\mathbf{1}\right\}  }c_{2}^{+}c_{2}^{{}}.$ Since the first term does not
contribute to the dynamics, we retain only the second term and normalize\ it
to the occupation factor. This means that site $1$ is empty while site $2$ and
sites at the particle reservoir are \textquotedblleft full\textquotedblright.
This describes the tiny excess above the mean occupation $1/2.$ To find the
dynamics of the reduced density matrix of the \textquotedblleft
system\textquotedblright\ $\mathbf{\sigma}\left(  t\right)  =\frac{\hbar
}{\mathrm{i}}\mathbf{G}^{<}\left(  t\right)  ,$ we will take advantage of the
particle representation and use an \textit{integral} form
\cite{Keldysh,Keldysh2} of the Keldysh formalism instead of the standard
Liouville-von Neumann differential equation. There, any perturbation term is
accounted to infinite order ensuring the proper density normalization. The
interaction with the bath is \textit{local} and\textit{, }because of the fast
dynamics in the bath, it can be taken as \textit{instantaneous.} Hence, the
evolution is further simplified into an integral form of the Generalized
Landauer B\"{u}ttiker Equation (GLBE) \cite{GLBE1,GLBE2} for the particle
density. There, the environment plays the role of a local measurement
apparatus. Firstly, we discuss the physics in this GLBE by developing a
discrete time version where the measurements occur at a \textit{stroboscopic
time} $\tau_{\mathrm{str.}}$.

\subsection{\textbf{Stroboscopic Decoherent Model}}

\begin{figure}
[th]
\begin{center}
\includegraphics[
height=4.20612in,
width=3.30534in
]%
{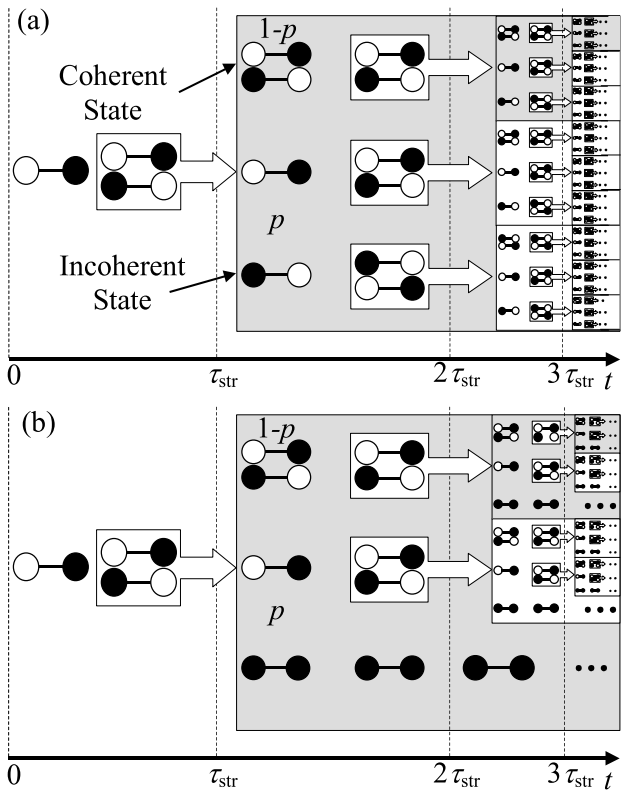}%
\caption{{}Quantum branching sequence for the swapping dynamics. Panel
(\textbf{a}) stands for an Ising system-environment interaction and
(\textbf{b}) an isotropic one. Single states represent states with interrupted
evolution (incoherent) while pairs of states are coherent superpositions.
Notice the self-similar structure.}%
\label{figstrobosc}%
\end{center}
\end{figure}

We introduce our computational procedure operationally for an Ising form
($\beta/\alpha=0$) of $\mathcal{V}$. The initial state of the isolated
two-spin system evolves with $\mathcal{H}_{\mathrm{S}}$. At $\tau
_{\mathrm{str.}}$, the spin bath interacts instantaneously with the
\textquotedblleft system\textquotedblright\ interrupting it with a probability
$p$. The actual physical time for the SE interaction is then obtained as
$\tau_{\mathrm{SE}}=\tau_{\mathrm{str.}}/p$. Considering that the dynamical
time scale of the bath ($\tau_{\mathrm{B}}\simeq\hbar/d_{\mathrm{B}}$) is much
faster than that of the system (fast fluctuation approximation), the dynamics
of site $3$ produces an energy fluctuation on site $2$ that destroys the
coherence of the two-spin \textquotedblleft system\textquotedblright.\ This
represents the \textquotedblleft measurement\textquotedblright\ process that
collapses the \textquotedblleft system\textquotedblright\ state. At
time\ $\tau_{\mathrm{str.}}$, the \textquotedblleft system\textquotedblright%
\ evolution splits into three alternatives: with probability $1-p$ the state
survives the interruption and continues its undisturbed evolution, while with
probability $p$ the system is effectively interrupted and its evolution starts
again from each of the \emph{two }eigenstates of $c_{2}^{+}c_{2}^{{}}$. These
three possible states at $\tau_{\mathrm{str.}}$ evolve freely until the system
is monitored again at time $2\tau_{\mathrm{str.}}$ and a new branching of
alternatives is produced as represented in the scheme of Fig.
\ref{figstrobosc}a.%

A similar reasoning holds when $\beta\neq0.$ The sequence for isotropic
interaction ($\alpha=\beta$) is shown in Fig. \ref{figstrobosc}b. The $XY$
part of $\mathcal{V}$ \ can inject a particle. When an interruption occurs,
the bath \textquotedblleft measures\textquotedblright\ at site $2$ and, if
found empty, it injects a particle. In the figure, this can be interpreted as
a \textquotedblleft pruning\textquotedblright\ of some incoherent branches
increasing the global coherence. This explains why the rate of\emph{
\textit{decoherence is greater when }}the\emph{\textit{ Ising}} part of
$\mathcal{V}$\emph{ dominates over the }$XY$\emph{ }part\emph{. }This occurs
with the dipolar interaction and, in less degree, with the isotropic one. This
contrasts with a pure $XY$ interaction where the survival of spin coherences
is manifested by a \textquotedblleft spin wave\textquotedblright%
\ behavior.\cite{Madi1997}\emph{ }The empty site of the bath is refilled and
de-correlates in a time much shorter than the swapping between sites $1$ and
$2$ (fast fluctuation approximation). Consequently, the injection can only
occur from the bath toward the system.

\subsection{\textbf{Analytical solution}}

The free evolution between interruptions is governed by the
system's\ propagators $\mathbf{U}_{\mathrm{S}}\left(  t\right)  =\exp
[-\mathrm{i}\mathcal{H}_{\mathrm{S}}t/\hbar]=-\frac{\hbar}{\mathrm{i}%
}\mathbf{G}^{0\mathrm{R}}\left(  t\right)  $. The spin bath interacts with the
system stroboscopically through an instantaneous \emph{interruption function
}$\widetilde{\mathbf{\Sigma}}^{<}\left(  t\right)  =\frac{1}{\mathrm{i}\hbar
}\frac{\tau_{\mathrm{str.}}}{p^{{}}}\mathbf{\Sigma}^{<}\left(  t\right)  $.
The reduced density function is%
\begin{gather}
\mathbf{\sigma}\left(  t\right)  =\mathbf{U}_{\mathrm{S}}\left(  t\right)
\mathbf{\sigma}\left(  0\right)  \mathbf{U}_{\mathrm{S}}^{\dag}\left(
t\right)  \left(  1-p\right)  ^{n}+\label{GdiscretaGLBE}\\%
{\textstyle\sum\limits_{m=1}^{n}}
\mathbf{U}_{\mathrm{S}}\left(  t-t_{m}\right)  \widetilde{\mathbf{\Sigma}}%
^{<}\left(  t_{m}\right)  \mathbf{U}_{\mathrm{S}}^{\dag}\left(  t-t_{m}%
\right)  p\left(  1-p\right)  ^{n-m},\nonumber
\end{gather}
where $n=\operatorname{int}\left(  t/\tau_{\mathrm{str.}}\right)  $ is the
number of interruptions and $t_{m}=m\tau_{\mathrm{str.}}$. The first term in
the rhs is the coherent system evolution weighted by its survival probability
$\left(  1-p\right)  ^{n}.$ This is the upper branch in Fig. \ref{figstrobosc}%
. The second term is the incoherent evolution involving all the decoherent
branches. The $m$-th term in the sum represents the evolution that had its
last interruption at $m\tau_{\mathrm{str.}}$ and survived coherently from
$m\tau_{\mathrm{str.}}$ to $n\tau_{\mathrm{str.}}$. Each of these is composed
by all the interrupted branches in Fig. \ref{figstrobosc}\textit{ }with a
single state at the hierarchy level $m$ with whom the paired state in the
upper branch, generated at time $\left(  m+1\right)  \tau_{\mathrm{str.}},$
keeps coherence up to $n\tau_{\mathrm{str.}}$.

In a continuous time process ($\tau_{\mathrm{SE}}=\tau_{\mathrm{str.}}/p$ with
$p\rightarrow0$ and $\tau_{\mathrm{str.}}\rightarrow0$), the above procedure
gives a physical meaning for the Keldysh's self-energy \cite{GLBE1,GLBE2} as
an \emph{instantaneous interruption function: }$\mathbf{\Sigma}^{<}\left(
t\right)  =\mathbf{\Sigma}_{\mathrm{m}}^{<}\left(  t\right)  +\mathbf{\Sigma
}_{\mathrm{i}}^{<}\left(  t\right)  $ (see appendix). Dissipation processes
are in $\mathbf{\Sigma}_{\mathrm{i}}^{<}\left(  t\right)  $ while
$\mathbf{\Sigma}_{\mathrm{m}}^{<}\left(  t\right)  $ involves only
decoherence. The term $\mathbf{\Sigma}_{\mathrm{i}}^{<}\left(  t\right)  $
injects a particle, increasing the system density, provided that site $2$ is
empty. This occurs at an \emph{injection} \emph{rate} $p_{\mathrm{XY}}%
/\tau_{\mathrm{SE}}=\Gamma_{\mathrm{XY}}/\hbar$ with $p_{\mathrm{XY}}=\left[
\beta^{2}/\left(  \alpha^{2}+\beta^{2}\right)  \right]  ,$ the $XY$
interaction weight. The decoherent part, $\mathbf{\Sigma}_{\mathrm{m}}%
^{<}\left(  t\right)  ,$ is a consequence of the \textquotedblleft
measurement\textquotedblright\ (\emph{or interruption}) performed by the
environment. It\ collapses the state vanishing the non-diagonal terms
(coherences) with a \emph{measurement rate} $\Gamma_{\mathrm{ZZ}}%
/\hbar=\left(  1-p_{\mathrm{XY}}\right)  /\tau_{\mathrm{SE}}.$ This also
occurs with the $XY$ interaction with rate $\Gamma_{\mathrm{XY}}/\hbar$
regardless of the fact that an effective injection takes place. Then, the
inverse of the survival time or \emph{interruption rate} is $\hbar
/\tau_{\mathrm{SE}}=\Gamma_{\mathrm{SE}}=\Gamma_{\mathrm{ZZ}}+\Gamma
_{\mathrm{XY}}$. Unlike $\mathbf{\Sigma}_{\mathrm{i}}^{<}\left(  t\right)  $,
this process conserves the total polarization. We denote by $\Gamma/\hbar$ the
rate associated to the isotropic interaction ($\alpha=\beta=1$). Then, the
anisotropic rates are: $\Gamma_{\mathrm{ZZ}}=\alpha^{2}\Gamma$ and
$\Gamma_{\mathrm{XY}}=\beta^{2}\Gamma,$ then $\hbar/\tau_{\mathrm{SE}}%
=\Gamma_{\mathrm{SE}}=\left(  \alpha^{2}+\beta^{2}\right)  \Gamma$. The rates
can be calculated to infinite order in perturbation theory from the solution
of the bath dynamics in a chain of spins with $XY$
interactions,\cite{condmat2004} which in the limit $d_{23}/d_{\mathrm{B}%
}\rightarrow0$ gives $\Gamma_{\mathrm{XY}}=2\pi\beta^{2}d_{23}^{2}(\pi
d_{\mathrm{B}})^{-1}=\beta^{2}\Gamma$ which recovers a Fermi golden rule evaluation.

The anisotropy ratio $\left(  \alpha/\beta\right)  ^{2}$ accounts for the
observed competition \cite{JCP2003} between the Ising and $XY$ terms of
$\mathcal{V}$. The Ising interaction drives the \textquotedblleft
system\textquotedblright\ to the internal quasi-equilibrium$.$ In contrast,
the $XY$ term allows the thermal equilibration with the bath.\cite{JCP2003}
Solving Eq. (\ref{GdiscretaGLBE}) in the limit of continuos time processes we
obtain our main analytical result for the swapping probability (experimental
$^{13}$C polarization):%
\begin{equation}
P\left(  t\right)  =1-a_{0}e^{-R_{0}t}-a_{1}\cos\left[  \left(  \omega
+\mathrm{i}R_{2}\right)  t+\phi_{0}\right]  e^{-R_{1}t}, \label{PolCST}%
\end{equation}
which, in spite of appearance, has a \textit{single fitting parameter}. This
is because the real functions $\omega,$ $R_{0},R_{1}$ and $R_{2}$ as well as
$a_{0}$, $a_{1}$ and $\phi_{0}$ depend exclusively on $b,$ $\tau_{\mathrm{SE}%
}$ and $p_{\mathrm{XY}}$. Besides, $b$ and $p_{\mathrm{XY}}$ are determined
from crystallography and the anisotropy of the magnetic interaction
($p_{\mathrm{XY}}=1/5$ for dipolar) respectively. The phase transition is
ensured by the condition $\omega R_{2}\equiv0$. The complete analytical
expression is given in the appendix.

\subsection{\textbf{Limiting cases}}

Typical solutions of the quantum master equation \cite{QME} for a spin
swapping \cite{JCP2003} follow that of MKBE \cite{MKBE}. They considered an
\textit{isotropic interaction} with the spin environment, represented\ by a
phenomenological relaxation rate $1/\left(  2\tau_{\mathrm{SE}}\right)  .$
Within the fast fluctuation approximation and neglecting non-secular terms,
this leads to $P^{^{\mathrm{MKBE}}}\left(  t\right)  ,$ used in most of the
experimental fittings. Our Eq. (\ref{PolCST}) reproduces this result with
$1/\left(  2\tau_{\phi}\right)  \equiv R\simeq1/\left(  2\tau_{\mathrm{SE}%
}\right)  $ by considering an isotropic interaction Hamiltonian $\alpha
=\beta=1$ under the condition $1/\left(  2\tau_{\mathrm{SE}}\right)  \ll
b/\hbar.$ However, at short times $t\ll\tau_{\mathrm{SE}},$ the MKBE swapping
probability growths exponentially with a rate $1/\left(  2\tau_{\mathrm{SE}%
}\right)  $. In contrast, our solution manifests the quadratic polarization
growth in time, $\left(  \frac{1}{2}b/\hbar\right)  ^{2}t^{2}.$ In Eq.
(\ref{PolCST}) this is made possible by the phase $\phi_{0}$ in the cosine. In
the opposite parametric region, $b\tau_{\mathrm{SE}}\ll\hbar$, \ our model
enables the manifestation of the quantum Zeno
effect.\cite{Usaj,Misra,PascazioQZSub2002} This means that the bath interrupts
the system through measurements too frequently, freezing its evolution. At
longer times, $t\gg\tau_{\mathrm{SE}}$, one gets $1-P\left(  t\right)
\sim\left(  1+\frac{1}{2}\left(  b/\hbar\right)  ^{2}\tau_{\mathrm{SE}}%
^{2}\right)  \exp\left[  -\frac{1}{2}(b/\hbar)^{2}\tau_{\mathrm{SE}}^{{}%
}t\right]  ,$ and the quantum Zeno effect is manifested in the reduction of
the decay rate $1/\left(  2\tau_{\phi}\right)  \propto\left(  b/\hbar\right)
^{2}\tau_{\mathrm{SE}}$ as $\tau_{\mathrm{SE}}$ gets smaller than $\hbar/b$.
This surprising dependence deserves some interpretation. First, we notice that
a strong interaction with the bath makes the $^{1}$H spin to fluctuate,
according to the Fermi golden rule, at a rate $1/\tau_{\mathrm{SE}}^{{}}$. The
effect on the $^{13}$C is again estimated in a fast fluctuation approximation
as $1/\tau_{\phi}\propto\left(  b/\hbar\right)  ^{2}\tau_{\mathrm{SE}}%
$~$\propto\left(  b/\hbar\right)  ^{2}\left[  \left(  d_{23}/\hbar\right)
^{2}\tau_{\mathrm{B}}\right]  ^{-1}$. This \textquotedblleft
nesting\textquotedblright\ of two Fermi golden rule rates is formally obtained
from a continued fractions evaluation of the self-energies
\cite{DAmato,condmat2004} involving an infinite order perturbation theory.
Another relevant result is that the frequency depends not only on $b$, but
also on $\tau_{\mathrm{SE}}.$ A remarkable difference between the quantum
master equation and our formulation concerns the final state. In the quantum
master equation $\sigma\left(  \infty\right)  $ must be hinted beforehand,
while here it is reached dynamically from the interaction with the spin bath.
Here, the reduced density, whose trace gives the system polarization, can
fluctuate until it reaches its equilibrium value.

\subsection{\textbf{Comparison with the experiments}}

In order to see how well our model reproduces the experimental behavior we
plot the $^{13}$C polarization with realistic parameters. Since the system is
dominated by the dipolar SE interaction,\cite{JCP2003} we take $\left\vert
\alpha/\beta\right\vert =2.$ We introduce $b$ with its angular dependence
according to Eq. (\ref{b}) and we select a constant value for $\tau
_{\mathrm{SE}}=0.275~\mathrm{ms}$ representative of the $b\gg\hbar
/\tau_{\mathrm{SE}}$ regime. Since in this work, we are only interested in the
qualitative aspects of the critical behavior of the dynamics, there is no need
to introduce $\tau_{\mathrm{SE}}(\theta)$ as a fitting parameter. These
evolutions are normalized at the maximum contact time ($3\ \mathrm{ms}$)
experimentally acquired. They are shown in Fig. \ref{Figteoexp3D}b where the
qualitative agreement with the experimental observation of a canyon is
evident. Notice that the experimental canyon is less deep than the theoretical
one. This is due to intermolecular $^{13}$C-$^{1}$H couplings neglected in the
model. We will show that the analytical expression of Eq. (\ref{PolCST})
allows one to determine the edges of the canyon which are the critical points
of what we will call a \emph{Quantum Dynamical Phase Transition}.

\section{QUANTUM\textbf{ DYNAMICAL PHASE TRANSITION}}

Our \textit{quantum} observable (the local spin polarization) is a binary
random variable. The dynamics of its ensemble average (swapping probability),
as described by\ Eq. (\ref{PolCST}), depends parametrically on the
\textquotedblleft noisy\textquotedblright\ fluctuations of the environment
through $\tau_{\mathrm{SE}}$. Thus, following\ Horsthemke and
Lefever,\cite{Horsthemke-Lefever} one can identify the precise value for
$\tau_{\mathrm{SE}}$ where a qualitative change in\ the functional form of
this probability occurs as the \textit{critical point of a phase transition.}
This is evidenced by the functional change (nonanalyticity) of the dependence
of the observables (e.g. the swapping frequency $\omega$) on the control
parameter $b\tau_{\mathrm{SE}}/\hbar$. Since\ the control parameter switches
among \textit{dynamical regimes} we call this phenomenon a \textit{Quantum
Dynamical Phase Transition. }

It should be remarked that the effect of other spins on the two spin system
introduces non-commuting perturbing operators (symmetry breaking
perturbations) which produce non-linear dependences of the observables. While
this could account for the limiting dynamical regimes, it does not ensure a
phase transition. A true phase transition needs a non-analyticity in these
functions which is only enabled by taking the thermodynamic limit of an
infinite number of spins.\cite{sachdev} In our formalism, this is incorporated
through the imaginary part of the energy, $\hbar/\tau_{\mathrm{SE}},$
evaluated from the Fermi Golden Rule.\cite{condmat2004}

When the SE interaction is \textit{anisotropic} ($\alpha\neq\beta$), there is
a functional dependence of $\omega\ $on $\tau_{\mathrm{SE}}$ and $b$ yielding
a critical value for their product, $b\tau_{\mathrm{SE}}/\hbar
=k_{p_{\mathrm{XY}}},$ where the dynamical regime switches. One identifies two
parametric regimes: 1- The \textit{swapping phase,} which is a form of
sub-damped dynamics, when $b\tau_{\mathrm{SE}}/k_{p_{\mathrm{XY}}}>\hbar$
($R_{2}=0$ in Eq. (\ref{PolCST})). 2- A \textit{Zeno phase}, with an
over-damped dynamics for $b\tau_{\mathrm{SE}}/k_{p_{\mathrm{XY}}}<\hbar$ as a
consequence of the strong coupling with the environment\ (\emph{zero
frequency}, i.e. $\omega=0$, in Eq. (\ref{PolCST})). In the neighborhood of
the critical point the swapping frequency takes the form (see appendix):%
\begin{equation}
\omega=\left\{
\begin{array}
[c]{cc}%
a_{p_{\mathrm{XY}}}^{{}}\sqrt{\left(  b/\hbar\right)  _{{}}^{2}%
-k_{p_{\mathrm{XY}}}^{2}/\tau_{\mathrm{SE}}^{2}} & b\tau_{\mathrm{SE}%
}/k_{p_{\mathrm{XY}}}>\hbar\\
0 & b\tau_{\mathrm{SE}}/k_{p_{\mathrm{XY}}}\leq\hbar
\end{array}
\right.  . \label{freq-crit}%
\end{equation}
The parameters $a_{p_{\mathrm{XY}}}$ and $k_{p_{\mathrm{XY}}}$ only depend on
$p_{\mathrm{XY}}$ which is determined by the origin of the interaction
Hamiltonian. For typical interaction Hamiltonians the values of these
parameters $\left(  p_{\mathrm{XY}},k_{p_{\mathrm{XY}}},a_{p_{\mathrm{XY}}%
}\right)  $ are: Ising $\left(  0,\frac{1}{2},1\right)  $, dipolar $\left(
\frac{1}{5},0.3564,0.8755\right)  $, isotropic $\left(  \frac{1}{2},0,\frac
{1}{\sqrt{2}}\right)  $ and $XY$ $\left(  1,1,1\right)  .$ The swapping period
is
\begin{equation}
T\simeq\frac{T_{0\mathrm{c}}^{3/2}}{\sqrt{2}a_{p_{\mathrm{XY}}}^{{}}}\ \left(
T_{0\mathrm{c}}-T_{0}\right)  ^{-1/2},
\end{equation}
where $T_{0}=\frac{2\pi\hbar}{b}$ is the isolated two spin period and its
critical value $T_{0\mathrm{c}}=\frac{2\pi\tau_{\mathrm{SE}}}%
{k_{p_{\mathrm{XY}}}},$ determines the region where the period $T$ diverges as
is observed in Fig. \ref{Figteoexp3D}. \ The estimated value of $\tau
_{\mathrm{SE}}=0.275~\mathrm{ms}$ and dipolar SE interactions yield a critical
value for the $^{13}$C-$^{1}$H coupling of $b_{\mathrm{c}}/\hbar
=2\pi/T_{0\mathrm{c}}=1.3~\mathrm{kHz.}$

\begin{figure}
[th]
\begin{center}
\includegraphics[
height=3.834in,
width=3.528in
]%
{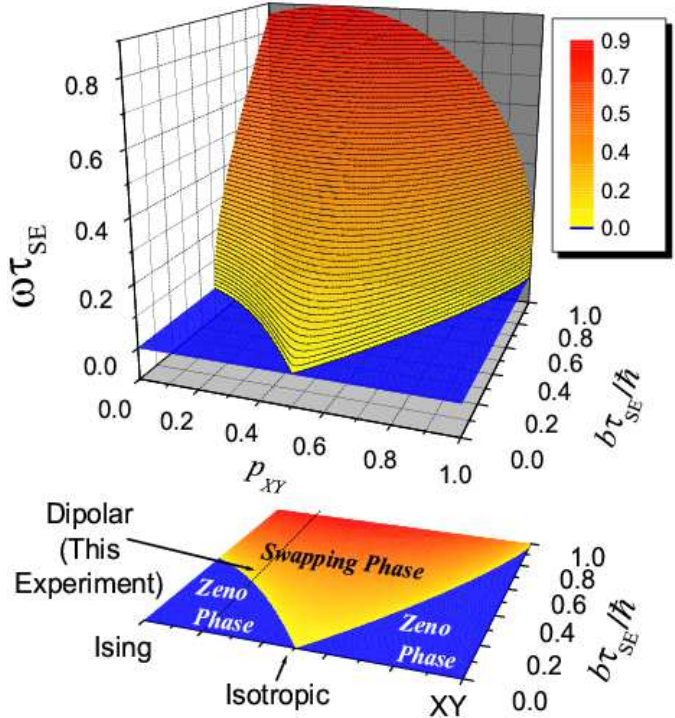}%
\caption{{}(Color online) Quantum dynamical phase diagram for the spin
swapping operation. The figure shows the frequency dependence on
system-environment (SE) interaction anisotropy $p_{\mathrm{XY}}$ and the ratio
among the internal and the SE interaction $b\tau_{\mathrm{SE}}/\hbar$. The
projection over the $b\tau_{\mathrm{SE}}/\hbar-p_{\mathrm{XY}}$ plane
determines the phase diagram where the transition between the swapping phase
into the Zeno phase ($\omega=0)$ is manifested. Values of $p_{\mathrm{XY}}$
for typical SE interaction Hamiltonian are indicated in the contour plot.}%
\label{phasediag}%
\end{center}
\end{figure}

The complete phase diagram that accounts for the anisotropy of the SE
interactions is shown in Fig. \ref{phasediag}. There, the frequency dependence
on $p_{\mathrm{XY}}$ and $b\tau_{\mathrm{SE}}/\hbar$ is displayed. At the
critical line the frequency becomes zero setting the limits between both
dynamical phases.%

The two dynamical phases can now be identified in the NMR experiments which up
to date defied explanation. The experimental setup provides a full scan of the
parameter $b\tau_{\mathrm{SE}}$ through the phase transition that is
manifested when the frequency goes suddenly to zero (Fig. \ref{Figexp}b) and
the relaxation rate (Fig. \ref{Figexp}a) changes its behavior decaying
abruptly. The fact that $1/\left(  2\tau_{\phi}\right)  $ tends to zero when
$b\ll\hbar/\tau_{\mathrm{SE}}$ confirms the Zeno phase predicted by our model.
In this regime, $1/\left(  2\tau_{\phi}\right)  $ is quadratic on $b$ as
prescribed. To make the comparison between the two panels of Fig.
\ref{Figteoexp3D}\ quantitative, we fit the predicted dynamics of Fig.
\ref{Figteoexp3D}b with $P^{^{\mathrm{MKBE}}}\left(  t\right)  ,$ following
the same procedure used to fit the experimental data. The solid line in Fig.
\ref{Figexp} show the fitting parameters $1/\left(  2\tau_{\phi}\right)  $ and
$\omega$ in excellent agreement with the experimental ones.

We point out that Eq. (\ref{eq-MKBE}) is used to fit both the experiments and
the theoretical prediction of Eq. (\ref{PolCST}) because it constitutes a
simple, thought imperfect, way to extract the \textquotedblleft
outputs\textquotedblright\ (oscillation frequency\ $\omega$ and a decoherence
time~$\tau_{\phi}$). While the systematic errors shift the actual critical
value of the control parameter,\ $b/\hbar$, from $1.3~\mathrm{kHz}$ to $2~\mathrm{kHz}%
$, Eq. (\ref{eq-MKBE}) yields a simplified way to \textquotedblleft
observe\textquotedblright\ the transition.

\section{\textbf{CONCLUSIONS}}

We found experimental evidence that environmental interactions can drive a
swapping gate through a \emph{Quantum Dynamical Phase Transition} towards an
over-damped or Zeno phase. The NMR experiments in spin swapping in a $^{13}%
$C-$^{1}$H system enable the identification and characterization of this phase
transition as function of the ratio $b\tau_{\mathrm{SE}}/\hbar$ between the
internal and SE interaction. We developed a model that describes both phases
and the critical region with great detail, showing that it depends only on the
nature of the interaction. In particular the phase transition does not occurs
if the SE interaction is restricted to isotropic spin coupling. The phase
transition is manifested not only in the observable swapping frequency but
also in the decoherence rate $1/\tau_{\phi}$. While a standard Fermi golden
rule perturbative estimation would tend to identify this rate with the SE
interaction, i.e. $1/\tau_{\phi}\cong1/\tau_{\mathrm{SE}}^{{}}$ $\simeq\left(
d_{23}^{{}}/\hbar\right)  ^{2}\tau_{\mathrm{B}},$ as it occurs well inside the
swapping phase, both rates differ substantially\ as the system enters the Zeno
phase ($b\tau_{\mathrm{SE}}^{{}}\leq k_{p_{\mathrm{XY}}}\hbar$). Here the
decoherence rate switches to the behavior $1/\tau_{\phi}^{{}}\propto\left(
b/\hbar\right)  _{{}}^{2}\tau_{\mathrm{SE}}^{{}}$. In the Zeno phase, the
system's free evolution decays very fast with a rate $\tau_{\mathrm{SE}}^{-1}%
$. In spite of this, one can see that the initial state as a whole has a slow
decay (its dynamics becomes almost frozen) because it is continuously fed by
the environment. Since the $\tau_{\mathrm{SE}}^{{}}$ has become the
correlation time for the spin directly coupled to the environment,
$1/\tau_{\phi}^{{}}$ provided by our calculation can be interpreted as a
\textquotedblleft nested\textquotedblright\ Fermi golden rule rate emphasizing
the non-perturbative nature of the result. Based on the wealth of this simple
swapping dynamics, we can foresee applications that range from tailoring the
environments for a reduction of their decoherence on a given process to using
the observed critical transition in frequency and decoherence rate as a tracer
of the environment's nature. These applications open new opportunities for
both, the field of quantum information processing and the general physics and
chemistry of open quantum systems.\cite{condmat2005}

\begin{acknowledgments}
This work was funded by Fundaci\'{o}n Antorchas, CONICET, ANPCyT, and
SeCyT-UNC. P.R.L. and H.M.P. are members of the Research Career. E.P.D. and
G.A.A. are Doctoral Fellows of CONICET. We are very grateful to Richard R.
Ernst, who drove our attention to the puzzling nature of our early
experiments. We benefited from discussions with Lucio Frydman, who suggested
experimental settings where these results are relevant, as well as stimulating
comments from Alex Pines during a stay of PRL and HMP at the Weizmann
Institute of Science. Boris Altshuler made suggestions on the different time
scales. Jorge L. D'Amato and Gonzalo Usaj had a creative involvement in the
very early stages of this project.
\end{acknowledgments}

\appendix

\section{}

\subsection{\textbf{Discrete time process}}

It is convenient to rewrite Eq. (\ref{GdiscretaGLBE}) within the Keldysh
formalism where $\mathbf{G}^{<}\left(  t\right)  =\frac{\mathrm{i}}{\hbar
}\mathbf{\sigma}\left(  t\right)  $ and the evolution operator for the
isolated system is $\mathbf{G}^{0\mathrm{R}}\left(  t\right)  =-\frac
{\mathrm{i}}{\hbar}\mathbf{U}_{\mathrm{S}}(t)$, where $\mathbf{G}%
^{0\mathrm{A}}\left(  t\right)  =\mathbf{G}^{0\mathrm{R}}\left(  t\right)
^{\dag}$, and the interruption function \emph{ }$\widetilde{\mathbf{\Sigma}%
}^{<}\left(  t\right)  =\frac{1}{\mathrm{i}\hbar}\frac{\tau_{\mathrm{str.}}%
}{p^{{}}}\mathbf{\Sigma}^{<}\left(  t\right)  .$ We rearrange expression
(\ref{GdiscretaGLBE}), for $n\tau_{\mathrm{str.}}<t\leq(n+1)\tau
_{\mathrm{str.}},$ in terms of $n\tau_{\mathrm{str.}},$ the last interruption
time:%
\begin{gather}
\tfrac{1}{\hbar^{2}}\mathbf{G}^{<}\left(  t\right)  =\mathbf{G}^{0\mathrm{R}%
}\left(  t-t_{n}\right)  \mathbf{G}^{<}\left(  t_{n}\right)  \mathbf{G}%
^{0\mathrm{A}}\left(  t-t_{n}\right)  \left(  1-p\right) \nonumber\\
+\tfrac{\mathrm{i}}{\hbar}\mathbf{G}^{0\mathrm{R}}\left(  t-t_{n}\right)
\widetilde{\mathbf{\Sigma}}^{<}\left(  t_{n}\right)  \mathbf{G}^{0\mathrm{A}%
}\left(  t-t_{n}\right)  p, \label{GDiscreta}%
\end{gather}
which takes advantage of the self similarity of the hierarchy levels and is
simple to iterate. It not only reproduces the results obtained from the
quantum theory of continuous irreversible processes \cite{GLBE1,GLBE2}
discussed below, but it provides a very efficient algorithm which reduces
memory storage and calculation time substantially.

\subsection{\textbf{Continuous time process}}

In the limit $p\rightarrow0$ and $\tau_{\mathrm{str.}}\rightarrow0$ with
constant ratio, one gets a continuous process where $\tau_{\mathrm{SE}}%
=\tau_{\mathrm{str.}}/p$ defines a survival time for the evolution of the
isolated two-spin \textquotedblleft system\textquotedblright. Eq.
(\ref{GdiscretaGLBE}) becomes%
\begin{gather}
\mathbf{G}^{<}\left(  t\right)  =\hbar^{2}\mathbf{G}^{0\mathrm{R}}\left(
t\right)  \mathbf{G}^{<}\left(  0\right)  \mathbf{G}^{0\mathrm{A}}\left(
t\right)  e^{-t/\tau_{\mathrm{SE}}}\nonumber\\
+\int_{0}^{t}\mathrm{d}t_{n}\mathbf{G}^{0\mathrm{R}}\left(  t-t_{n}\right)
\mathbf{\Sigma}^{<}\left(  t_{n}\right)  \mathbf{G}^{0\mathrm{A}}\left(
t-t_{n}\right)  e^{-\left(  t-t_{n}\right)  /\tau_{\mathrm{SE}}}%
,\label{Gcontinua}%
\end{gather}
a new form of the GLBE \cite{GLBE1,GLBE2} that includes correlations
(non-diagonal terms in $\mathbf{G}^{<}$ and $\mathbf{\Sigma}^{<}$). Moreover,
keeping $\tau_{\mathrm{SE}}$ constant, under the conditions of finite
$\tau_{\mathrm{str.}}\lesssim\hbar/d_{23}$\emph{ }and\emph{ }$p\lesssim
d_{23}/d_{\mathrm{B}}$\emph{ }(here $d_{\mathrm{B}}$ is a mean intra-bath
interaction) there is no significant difference between the stroboscopic
approximation and the continuous time description. Thus, the physics of a
quantum theory of irreversible processes \cite{frerichs91} is obtained by
adapting a collapse theory \cite{Itano90} into the probabilistic branching
scheme represented in Fig. \ref{figstrobosc}. The solutions of Eqs.
(\ref{GdiscretaGLBE}) and\ (\ref{Gcontinua}) are both computationally
demanding since they involve a storage of all the previous time steps and
reiterated summations. However, Eq. (\ref{GDiscreta}) provides a \textit{new
computation procedure} that only requires the information of a \textit{single}
previous step. Besides, it avoids resorting to the random averages required by
models that include decoherence through stochastic or kicked-like
perturbations.\cite{paz2003,molmer1992} Thus, it becomes a very practical tool
to compute the dynamics in presence of decoherent and dissipation processes.\ 

The \emph{instantaneous interruption function: }$\mathbf{\Sigma}^{<}\left(
t\right)  =\mathbf{\Sigma}_{\mathrm{m}}^{<}\left(  t\right)  +\mathbf{\Sigma
}_{\mathrm{i}}^{<}\left(  t\right)  $ becomes, in matrix form,
\begin{align}
\mathbf{\Sigma}_{\mathrm{m}}^{<}\left(  t\right)   &  =\mathrm{i}\tfrac{\hbar
}{\tau_{\mathrm{SE}}}\left(
\begin{array}
[c]{cc}%
\tfrac{\hbar}{\mathrm{i}}G_{11}^{<}\left(  t\right)  & 0\\
0 & \tfrac{\hbar}{\mathrm{i}}G_{22}^{<}\left(  t\right)
\end{array}
\right)  ,\label{Sigmam}\\
\mathbf{\Sigma}_{\mathrm{i}}^{<}\left(  t\right)   &  =2\mathrm{i}\tfrac
{\hbar~p_{\mathrm{XY}}}{\tau_{\mathrm{SE}}}\left(
\begin{array}
[c]{cc}%
0 & 0\\
0 & \left[  1-\tfrac{\hbar}{\mathrm{i}}G_{22}^{<}\left(  t\right)  \right]
\end{array}
\right)  , \label{Sigmai}%
\end{align}
where the physical meaning of the Keldysh's self-energy \cite{GLBE1,GLBE2}
becomes evident.

Inserting Eq. (\ref{Sigmam}) and (\ref{Sigmai}) in Eq. (\ref{Gcontinua}) we
obtain our main analytical result, Eq. (\ref{PolCST}),%

\begin{align}
P\left(  t\right)    & =\tfrac{\hbar}{\mathrm{i}}G_{11}^{<}\left(  t\right)
\nonumber\\
& =1-a_{0}e^{-R_{0}t}-a_{1}\cos\left[  \left(  \omega+\mathrm{i}R_{2}\right)
t+\phi_{0}\right]  e^{-R_{1}t},
\end{align}
here all parameters are real with%
\begin{equation}
\omega+\mathrm{i}R_{2}=\frac{\sqrt{3}}{2x}\left(  \frac{1}{6}\eta\left(
p_{\mathrm{XY}},x\right)  +6\frac{\phi\left(  p_{\mathrm{XY}},x\right)  }%
{\eta\left(  p_{\mathrm{XY}},x\right)  }\right)  b,\label{omega}%
\end{equation}
where $\omega R_{2}\equiv0$ and are evaluated with $x=b\tau_{\mathrm{SE}%
}/\hbar$ using%
\[
\phi\left(  p_{\mathrm{XY}},x\right)  =\frac{1}{3}\left(  x^{2}-p_{\mathrm{XY}%
}^{2}-\frac{1}{3}\left(  1-p_{\mathrm{XY}}\right)  ^{2}\right)  ,
\]
and%
\begin{multline*}
\eta\left(  p_{\mathrm{XY}},x\right)  =\\
\left\{  4\left(  1-p_{\mathrm{XY}}\right)  \left(  9x^{2}-2\left(
1-p_{\mathrm{XY}}\right)  ^{2}+18p_{\mathrm{XY}}^{2}\right)  \right.  \\
\left.  +12\left[  3\left(  4x^{6}-\left(  \left(  1-p_{\mathrm{XY}}\right)
^{2}+12p_{\mathrm{XY}}^{2}\right)  x^{4}+\right.  \right.  \right.  \\
\left.  \left.  \left.  4p_{\mathrm{XY}}^{2}\left(  5\left(  1-p_{\mathrm{XY}%
}\right)  ^{2}+3p_{\mathrm{XY}}^{2}\right)  x^{2}\right.  \right.  \right.  \\
\left.  \left.  \left.  -4p_{\mathrm{XY}}^{2}\left(  \left(  1-p_{\mathrm{XY}%
}\right)  ^{2}-p_{\mathrm{XY}}^{2}\right)  ^{2}\right)  \right]  ^{\frac{1}%
{2}}\right\}  ^{\frac{1}{3}}.
\end{multline*}
Also,
\begin{multline}
R_{0}=\left(  6\frac{\phi\left(  p_{\mathrm{XY}},x\right)  }{\eta\left(
p_{\mathrm{XY}},x\right)  }-\frac{1}{6}\eta\left(  p_{\mathrm{XY}},x\right)
+\right.  \\
\left.  p_{\mathrm{XY}}+\frac{1}{3}\left(  1-p_{\mathrm{XY}}\right)  \right)
\frac{1}{\tau_{\mathrm{SE}}},
\end{multline}%
\begin{equation}
R_{1}=\frac{3}{2}\left(  p_{\mathrm{XY}}+\frac{1}{3}\left(  1-p_{\mathrm{XY}%
}\right)  \right)  \frac{1}{\tau_{\mathrm{SE}}}-\frac{R_{0}}{2},
\end{equation}
and
\begin{align*}
a_{0} &  =\frac{1}{2}\frac{2\left(  \omega_{{}}^{2}-R_{2}^{2}\right)
+2R_{1}^{2}-b_{{}}^{2}}{\left(  \omega_{{}}^{2}-R_{2}^{2}\right)  +\left(
R_{0}^{{}}-R_{1}^{{}}\right)  ^{2}},\\
a_{2} &  =\frac{1}{2\left(  \omega_{{}}^{{}}+\mathrm{i}R_{2}^{{}}\right)
}\times\\
&  \frac{\left(  2R_{0}^{{}}R_{1}^{{}}-b_{{}}^{2}\right)  \left(  R_{0}%
-R_{1}\right)  +2\left(  \omega_{{}}^{2}-R_{2}^{2}\right)  R_{0}^{{}}}{\left(
\omega_{{}}^{2}-R_{2}^{2}\right)  +\left(  R_{0}^{{}}-R_{1}^{{}}\right)  ^{2}%
},\\
a_{3} &  =\frac{1}{2}\frac{b_{{}}^{2}+2R_{0}^{2}-4R_{0}^{{}}R_{1}^{{}}%
}{\left(  \omega_{{}}^{2}-R_{2}^{2}\right)  +\left(  R_{0}^{{}}-R_{1}^{{}%
}\right)  ^{2}},\\
a_{1}^{2} &  =a_{2}^{2}+a_{3}^{2},\;\;\;\;\;\tan\left(  \phi_{0}\right)
=-\frac{a_{2}}{a_{3}}.
\end{align*}

The parametric dependence of the swapping frequency $\omega$, Eq.
(\ref{omega}), has a critical point at $x=k_{p_{\mathrm{XY}}},$ where the
transition occurs. In the neighborhood of this critical point $\omega$ and
$R_{2}$ takes the form:%
\begin{equation}
\omega=\left\{
\begin{array}
[c]{cc}%
a_{p_{\mathrm{XY}}}^{{}}\sqrt{\left(  b/\hbar\right)  _{{}}^{2}%
-k_{p_{\mathrm{XY}}}^{2}/\tau_{\mathrm{SE}_{{}}}^{2}} & b\tau_{\mathrm{SE}%
}/k_{p_{\mathrm{XY}}}>\hbar\\
0 & b\tau_{\mathrm{SE}}/k_{p_{\mathrm{XY}}}\leq\hbar
\end{array}
\right.  ,
\end{equation}
and%
\[
R_{2}=\left\{
\begin{array}
[c]{cc}%
0 & b\tau_{\mathrm{SE}}/k_{p_{\mathrm{XY}}}>\hbar\\
a_{p_{\mathrm{XY}}}^{{}}\sqrt{k_{p_{\mathrm{XY}}}^{2}/\tau_{\mathrm{SE}_{{}}%
}^{2}-\left(  b/\hbar\right)  _{{}}^{2}} & b\tau_{\mathrm{SE}}%
/k_{p_{\mathrm{XY}}}\leq\hbar
\end{array}
\right.  ,
\]
where%

\begin{multline}
k_{p_{\mathrm{XY}}}^{2}=\frac{1}{12}\left\{  \left[  \left(  p_{\mathrm{XY}%
}-1\right)  ^{2}\chi\left(  p_{\mathrm{XY}}\right)  \right]  ^{\frac{1}{3}%
}+\zeta\left(  p_{\mathrm{XY}}\right)  +\right.  \\
\left.  19p_{\mathrm{XY}}^{2}+\frac{\left(  p_{\mathrm{XY}}-1\right)
^{\frac{4}{3}}\zeta\left(  p_{\mathrm{XY}}\right)  }{\left[  \chi\left(
p_{\mathrm{XY}}\right)  \right]  ^{\frac{1}{3}}}\right\}  ,
\end{multline}%
\begin{multline*}
\chi\left(  p_{\mathrm{XY}}\right)  =-5291p_{\mathrm{XY}}^{4}%
-1084p_{\mathrm{XY}}^{3}+546p_{\mathrm{XY}}^{2}-4p_{\mathrm{XY}}+1+\\
24\sqrt{3}p_{\mathrm{XY}}\sqrt{\left(  28p_{\mathrm{XY}}^{2}-2p_{\mathrm{XY}%
}+1\right)  ^{3}},
\end{multline*}%
\[
\zeta\left(  p_{\mathrm{XY}}\right)  =-215p_{\mathrm{XY}}^{2}-2p_{\mathrm{XY}%
}+1,
\]
and%
\begin{equation}
a_{p_{\mathrm{XY}}}^{2}=\frac{1}{2}\frac{\left(  f_{1}^{~~2/3}+36f_{2}\right)
\left(  -f_{3}f_{1}^{\ \ 2/3}+36f_{2}f_{3}+f_{1}f_{4}\right)  }{f_{1}%
^{~~5/3}f_{4}},\label{orden1}%
\end{equation}%
\begin{multline*}
f_{1}=36k_{p_{\mathrm{XY}}}-8+24p_{\mathrm{XY}}+48p_{\mathrm{XY}}^{2}\\
-36k_{p_{\mathrm{XY}}}p_{\mathrm{XY}}-64p_{\mathrm{XY}}^{3}+12f_{4},
\end{multline*}%
\[
f_{2}=1/3k_{p_{\mathrm{XY}}}-1/3p_{\mathrm{XY}}^{2}-1/9\left(
1-p_{\mathrm{XY}}\right)  ^{2},
\]%
\begin{align*}
f_{3}  & =-f_{4}+p_{\mathrm{XY}}f_{4}-6k_{p_{\mathrm{XY}}}-16p_{\mathrm{XY}%
}^{4}+20p_{\mathrm{XY}}^{3}\\
& -10p_{\mathrm{XY}}^{2}+13p_{\mathrm{XY}}^{2}k_{p_{\mathrm{XY}}%
}-2k_{p_{\mathrm{XY}}}p_{\mathrm{XY}}+k_{p_{\mathrm{XY}}},
\end{align*}%
\begin{multline*}
f_{4}=\left(  12k_{p_{\mathrm{XY}}}+96p_{\mathrm{XY}}^{4}k_{p_{\mathrm{XY}}%
}-120p_{\mathrm{XY}}^{3}k_{p_{\mathrm{XY}}}+\right.  \\
\left.  60p_{\mathrm{XY}}^{2}k_{p_{\mathrm{XY}}}-39k_{p_{\mathrm{XY}}%
}p_{\mathrm{XY}}^{2}+6k_{p_{\mathrm{XY}}}p_{\mathrm{XY}}-3k_{p_{\mathrm{XY}}%
}\right.  \\
\left.  -48p_{\mathrm{XY}}^{4}+48p_{\mathrm{XY}}^{3}-12p_{\mathrm{XY}}%
^{2}\right)  ^{\frac{1}{2}}.
\end{multline*}
In equation (\ref{orden1}) the functions $f_{1},$ $f_{2},$ $f_{3}$ and $f_{4}$
only depend of $p_{\mathrm{XY}}.$

\end{document}